\documentclass[useAMS]{mn2e}
\usepackage{psfig}
 \usepackage{url}
\usepackage{times}
\def\spose#1{\hbox to 0pt{#1\hss}}
\newcommand\lsim{\mathrel{\spose{\lower 3pt\hbox{$\mathchar"218$}}
     \raise 2.0pt\hbox{$\mathchar"13C$}}}
\newcommand\gsim{\mathrel{\spose{\lower 3pt\hbox{$\mathchar"218$}}
     \raise 2.0pt\hbox{$\mathchar"13E$}}}
\def\sc{Schwarzschild}
\def\ltsima{$\; \buildrel < \over \sim \;$}
\def\lsim{\lower.5ex\hbox{\ltsima}}
\def\gtsima{$\; \buildrel > \over \sim \;$}
\def\gsim{\lower.5ex\hbox{\gtsima}}

\title[Synchrotron masers and fast radio bursts]
{ Synchrotron masers and fast radio bursts}
\author[G. Ghisellini]
{G. Ghisellini$^1$  \thanks{E--mail: gabriele.ghisellini@brera.inaf.it}
  \\  \\
$^1$ INAF -- Osservatorio Astronomico di Brera, via E. Bianchi 46, I--23807 Merate, Italy \\
}

\begin{document}

\pagerange{\pageref{firstpage}--\pageref{lastpage}} \pubyear{2012}

\maketitle
\label{firstpage}

\begin{abstract}
Fast Radio Bursts (FRBs), with a typical duration of 1 ms and 1 Jy flux density
at GHz frequencies, have brightness temperatures exceeding $10^{33} K$, requiring 
a coherent emission process.
This can be achieved by bunching particles in volumes smaller than the
typical wavelength, but this may be challenging.
Alternatively, we can have maser emission. 
Under certain conditions, the synchrotron stimulated emission process can
be more important than true absorption, and a synchrotron maser can be created.
This occurs when the emitting electrons have a very narrow distribution
of pitch angles and energies.
This process overcomes the difficulties of having extremely dense bunches of particles 
and relaxes the light crossing time limits, since there is no simple relation between the
actual size of the source and the observed variability timescale.
\end{abstract}

\section{Introduction}

Fast radio bursts (FRBs) are transient sources of radio emission in the GHz band, lasting for 
a few ms, observed at high Galactic latitudes. 
The observed large dispersion measure  (DM=100--2000 pc cm$^{-3}$, larger than the Galactic value)
(Lorimer et al. 2007; Thornton et al. 2013, see also the review by Katz 2016c)
suggests an extragalactic origin, or, alternatively, a dense material close
to a Galactic source.
The radio flux can reach the Jy level, and the corresponding 
brightness temperature is
\begin{equation}
T_{\rm  B}\, =\, {F_\nu \, c^ 2 d^2_{\rm A}\over 2\pi k \nu^2 (c \Delta t)^2 }  
\, \sim \, 10^{34} \,  { F_{\nu, \rm Jy}\,  d^2_{\rm A, \, Gpc} 
\over \Delta t^2_{\rm ms} \, \nu^2_{\rm GHz} }\,\, {\rm K}
\label{tb}
\end{equation}
where $F_\nu$ is the K--corrected flux density at the frequency $\nu$, $d_{\rm A}$ is the angular distance
and $\Delta t_{\rm ms}$ is the observed duration timescale in milliseconds.
If FRBs are Galactic events, then the typical distance can be taken as 1 kpc, and 
$T_{\rm B}$ is a factor $10^{12}$ smaller.
In any case the derived huge values require a coherent emission.

The energy released is of the order of
\begin{equation}
E_{\rm FRB} \sim \nu L_\nu \Delta t  {\Delta\Omega\over 4\pi}   
\sim  10^{39}     \nu_{\rm GHz}  F_{\nu\, \rm Jy}  d^2_{\rm L, \, Gpc} \Delta t_{\rm ms}  
{\Delta\Omega \over 4\pi} \, {\rm erg}
\label{e}
\end{equation}
where $L_\nu$ is the monochromatic FRB luminosity and
$\Delta\Omega/ 4\pi $ accounts for collimation of the produced radiation into
a solid angle $\Delta\Omega$ (but not due to relativistic beaming).
If they are Galactic, the energy $E_{\rm FRB}\sim 10^{27} d_{\rm kpc}^2$ erg.
If the source is in relativistic motion with a speed $\beta c$ at an angle $\theta$ from
out line of sight, we can introduce the relativistic beaming factor 
$\delta\equiv 1/[\Gamma(1-\beta\cos\theta)]$ and find that the observed energy is:
\begin{equation}
E_{\rm FRB}\, =\, \delta^3 \nu^\prime L^\prime_{\nu^\prime} \Delta t^\prime  \, =\,   \delta^3 E_{\rm FRB}^\prime 
\end{equation}
where primes quantities are measured in the comoving frame, where the emission is assumed isotropic
(and thus $\Delta\Omega^\prime/4\pi =1$).

If extragalactic, FRBs should be associated to a host galaxy, and indeed
Keane et al. (2016) claimed to have found the host galaxy of 
FRB 150418, at $z = 0.492$, but this claim was first challenged by Williams \& Berger (2016), that 
pointed out that the  radio flux variability, thought to be produced by the same
source originating FRB 150418, was instead due to a background AGN.

Repetition was seen in one case (FRB 121102; Spitler et al. 2016; Scholz et al. 2016), 
demonstrating that the source is not destroyed by the energetic events that cause the bursts. 
This should exclude gamma--ray bursts (GRBs) as progenitors of FRBs.
Up to now, no counterpart was successfully associated to any FRBs, at any frequency.
Attempts were made especially for the repeating FRB 121102 
(Sholz et al. 2016) and FRB 140514 (Petroff et al. 2015) with no results.
This suggests that FRBs do not originate in nearby ($z < 0.3$) supernova remnants (Petroff et al. 2015). 
The FRB rate is still uncertain, but it is between $10^3$ and $10^4$ events per day (e.g. Champion et al. 2016).

Since we do not know for sure the distance of GRBs, hence their real power, 
many models have been suggested to explain their properties. 
We are living again what happened for GRBs before the
discovery of the first redshift.
If they are extragalactic, the energy released (Eq. \ref{e})
is too large for star flares, and we must invoke compact objects,
and probably relativistic beaming or at least some collimation, that would 
reduce the energetics at the expense to enhance the event rate by $4\pi/\Delta \Omega$.
Repetition excludes irreversible catastrophic events as progenitors, such as GRBs
or merging binaries, but magnetars are viable.
A number of possible progenitors have been suggested in the recent past,
including Soft Gamma Ray Repeaters (i.e. magnetars; Pen \& Connor 2015; Kulkarni et al. 2015;
Popov \& Postnov 2007, 2013; Lyubarski 2014; Katz 2016b),
giant pulses from pulsars (Katz 2016a; Keane et al. 2012, Cordes \& Wasserman 2015;
Connor e al. 2015; Lyutikov et al. 2016),
interaction of pulsars with a planet (Mottez \& Zarka 2014) or with asteroids or comets 
(Geng \& Huang 2015; Dai et al. 2016).

Most models assumes that FRBs are extragalactic, although not all assume cosmological
(i.e. redshift $z\sim$1) distances.
Alternatively, Loeb, Shvartzvald \& Maoz (2014) and Maoz et al. (2015) 
have proposed star flares (in our Galaxy) as progenitors (but see Kulkarni et al. 2014).
In these models the large dispersion measure is associated with the stellar corona material.

A problem that all models have to face is how to produce the observed brightness temperatures. 
All models invoke bunching of particles that emit coherently.
This in turn requires that  each bunch is contained in a region of size comparable
to the observed wavelength, namely a few cm.
In this Letter, I will argue that the problem  to have such bunches  
can be by--passed by the other way to produce coherent emission, namely a maser.
I will show that it is possible to have synchrotron masers as long as a few conditions are met.
Instead to focus on a specific model (for this, see Lyubarski 2014), I will search for the 
general conditions required to have a synchrotron maser operating at the observed radio frequencies.

\section{Synchrotron cross section}

Ghisellini \& Svensson (1991; hereafter GS91) derived the synchrotron cross
section making use of the Einstein coefficient for
spontaneous emission, stimulated emission, and true absorption.
They pointed out that although the total cross section is always positive,
it can be negative for absorption  angles $\psi$ greater than 
the typical emission angle $1/\gamma$.
Here $\psi$ is the angle between the direction of the incoming photon 
and the velocity of the electron, and $\gamma$ is the Lorentz factor of the electron.
They considered a system with {\it three energy levels}: initially, the electron
is at the intermediate level, (level 2, energy $\gamma m_{\rm e} c^2$) and can jump to level 
1 (energy $\gamma m_{\rm e} c^2 -h\nu$) by spontaneously emitting a photon.
It can respond to the arrival of a photon of energy $h\nu$ in two ways:
it can absorb it, jumping to level 3 (true absorption: energy $\gamma m_{\rm e} c^2 +h\nu$),
or it can be stimulated to emit another photon with the same phase, frequency and direction of the
incoming one (stimulated emission). 
In this case the electron jumps to level 1. 
Since both  true absorption and  stimulated emission are proportional
to the incoming radiation, it is customary to calculate the {\it net} absorption 
by making the difference of the two processes (e.g. when calculating the absorption
coefficient).

Setting $\epsilon\equiv h\nu/(m_{\rm e}c^2)$, measuring the electron energy
in units of $m_{\rm e}c^2$, and momentum $p$ in units of $m_{\rm e}c$, 
consider the electron at the initial energy level $\gamma_2$, and the
other two energy levels $\gamma_1 = \gamma_2-\epsilon$ and $\gamma_3=\gamma_2+\epsilon$.
In this case the Einstein coefficients are related  by
\begin{equation}
B_{21}={c^2 \over 2h\nu^3} A_{21}; \quad B_{23} = B_{32} =    {c^2 \over 2h\nu^3} A_{32}
\end{equation}
The emissivity for the single electron (in erg s$^{-1}$ Hz$^{-1}$ ster$^{-1}$) 
is related to the  Einstein coefficients by:
\begin{eqnarray}
j(\nu, \gamma_2, \psi) \,  &=&  \, 4\pi h\nu \gamma_1 p_1 A_{21} \nonumber \\
j(\nu, \gamma_2+\epsilon, \psi) \, &=& \, 4\pi h\nu \gamma_2 p_2 A_{32}
\end{eqnarray}
Where $\gamma_1 p_1$ and  $\gamma_1 p_1$ are the terms associated to the phase space. 
Note that the emissivity depends upon the phase space of ``arrival"
(namely the one corresponding to the energy of the electron after the transition).

The differential cross section for true absorption ($d\sigma_{\rm ta/d\Omega}$) 
and stimulated emission  ($d\sigma_{\rm se/d\Omega}$) can then be
written as (see GS91):
\begin{equation}
{d\sigma_{\rm ta} \over d\Omega} \, =\, 4\pi h \nu\gamma_3 p_3 B_{23} 
\,=\,  {c^2 \over 2h\nu^3} {\gamma_3 p_3\over \gamma_2 p_2} j(\nu, \gamma_2+\epsilon, \psi)
\label{ta}
\end{equation}
\begin{equation}
{d\sigma_{\rm se} \over d\Omega} \, =\, 4\pi h \nu \gamma_1 p_1 B_{21}
\,=\,  {c^2 \over 2h\nu^3}  j(\nu, \gamma_2, \psi)
\label{se}
\end{equation}
Note the following:
\begin{itemize}
\item 
Both cross sections are very large, of the order of
$\lambda^2 [j(\nu)/h\nu]$, namely the number of photons produced per unit time
by the electron multiplied by the square of their wavelength.

\item 
As can be seen, $\sigma_{\rm ta}$ and $\sigma_{\rm se}$ are almost equal.
The net cross section, that is going to define the absorption coefficient,
is the difference $\sigma_{\rm s} = \sigma_{\rm ta} -\sigma_{\rm se}$.

\item
The differential cross sections in Eq. \ref{ta} and Eq. \ref{se} refer to one particular
direction $\psi$ of the incoming photon.
In general, if the energy of the particle increases, so does its emissivity.
In addition, the phase space factor also  increases.
This is why, in general, the total cross section is positive.

\item 
On the other hand, there are special cases where the emissivity for
specific directions {\it decreases} when the particle energy is increased.
This occurs when the incoming photon arrives at an angle larger than
the characteristic beaming angle $1/\gamma$.
In this case ($\psi > 1/\gamma$) the increase of $\gamma$ makes $j(\nu, \gamma,\psi)$ to decrease,
possibly even more than the increase of the phase space factor.
In this case the stimulated emission is larger than the true absorption, the total
cross section becomes formally negative, and there is the possibility to have a maser or laser.

\end{itemize}
Following GS91 (see also Schwinger 1949; Jackson 1975; Rybicki \& Lightman 1979) 
we report here the single particle emissivity as a function of $\psi$.
First let us introduce the notation:
\begin{eqnarray}
\nu_{\rm L} &\equiv& {eB \over 2 \pi m_{\rm e} c  }; \quad 
\nu_{\rm c} \equiv  {3\over 2} \gamma^2 \nu_{\rm L}\sin\theta; \quad x\equiv  {\nu \over \nu_{\rm c}} \nonumber \\
t &\equiv& \psi^2\gamma^2;
\quad y\equiv {x \over 2 } (1+t)^{3/2} = {\nu \over 2 \nu_{\rm c}} (1+t)^{3/2}
\end{eqnarray}
Then we have
\begin{equation}
j(\nu, \gamma,\theta,\psi)  = {9 \sigma_{\rm T} c B^2 \over 8\pi^3 \nu_{\rm L}} \gamma x^2 (1+t) 
\left[ (1+t) K^2_{2/3}(y) +t K^2_{1/3}  \right]
\end{equation}
valid for $\gamma\gg 1$ and $\psi\ll 1$. Here $U_{\rm B}=B^2/(8\pi)$ is the magnetic energy density
and $\sigma_{\rm T}\sim 6.65\times 10^{-25}$cm$^{-2}$ is the Thomson cross section.
$K_a(y)$ is the modified Bessel function of order $a$.

Finally, the total differential cross section is
\begin{eqnarray}
{d\sigma_{\rm s} \over d\Omega} \, &=&\, 
 {2 \over 9} {e\over B} {1\over \gamma^4 \sin^2\theta}
\{ (13 t-11)(1+t) K^2_{2/3}(y)  \nonumber \\
 &+ &  t(11t-1)K^2_{1/3}(y) -6y(t-2)  \nonumber \\
 &\times& [ (1+t) K_{2/3}(y)K_{5/3}(y) +t K_{1/3}(y)K_{4/3}(y) ] \}  
 \label{ds}
 \end{eqnarray}
%

\begin{figure} 
\vskip -0.6 cm
\hskip -0.7 cm
\psfig{file=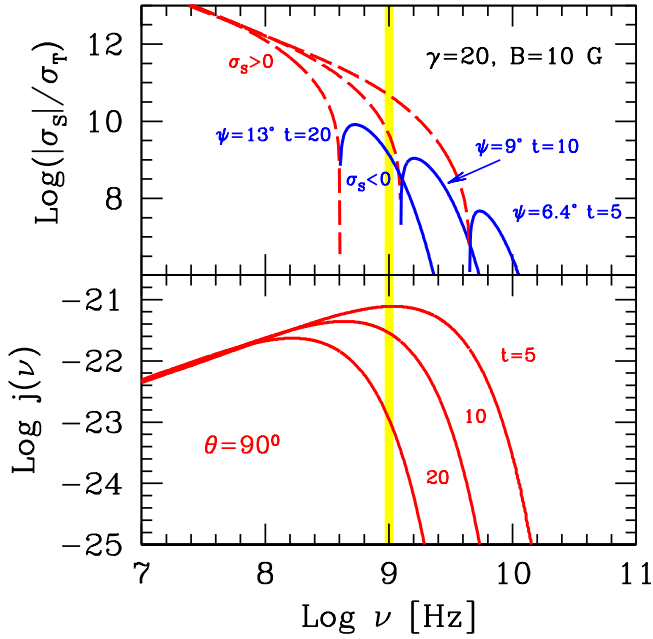,height=9.8cm} 
\vskip -0.5 cm
\caption{
Absorption cross section (top panel)
and single electron emissivity (bottom panel) as
a function of frequency, for $\gamma=20$ and $B=10$ G.
The three curves correspond to three values of $t\equiv\gamma^2\psi^2$,
hence for three values of the absorption angle $\psi$ (since $\gamma$ is fixed).
The dashed (red) lines correspond to the value of the positive branch of
the cross section, while the solid (blue) lines correspond to the 
modulus of the negative branch of $d\sigma /d\Omega$.
For all cases the pitch angle is $\theta=90^\circ$.
} 
\label{sigma}
\end{figure}

Fig. \ref{sigma} shows the absorption cross section (top panel)
and the single electron emissivity (bottom panel) of a particle
with $\gamma=20$ and a magnetic field $B=10$ G and for three
values of $t$, corresponding to absorption angles $\psi=6.4^\circ=2.2/\gamma$,
$9^\circ=3.2/\gamma$ and $13^\circ=4.5/\gamma$.
The dashed (red) lines correspond to the positive part of the cross section,
while the solid (blue) parts are the moduli of the negative part of the cross section.
We can see that for the chosen parameters the synchrotron absorption cross section
is orders of magnitude greater that the scattering Thomson $\sigma_{\rm T}$, and
that the relevant frequencies are in the radio band (the vertical yellow stripe
indicates 1 GHz).

\begin{figure} 
\vskip -0.6 cm
\hskip -0.7 cm
\psfig{file=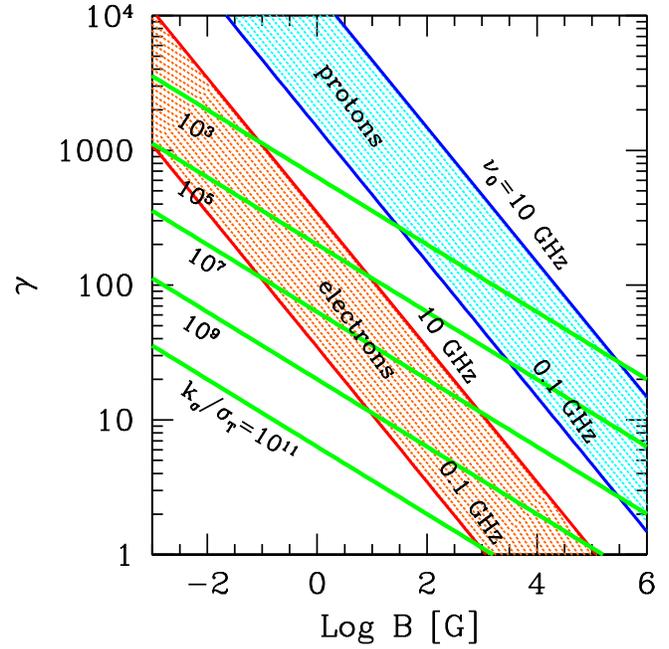,height=9.8cm} 
\vskip -0.5 cm
\caption{
The two general constraints to have efficient synchrotron masers 
are illustrated in the plane $\gamma$--$B$.
The (green) lines corresponds to different values of $k_\sigma$
(Eq. \ref{ksigma}), while the two hatched regions corresponds
to have the frequencies, at which the cross section turns 
negative, in the GHz band (Eq. \ref{bg}).
The hatched region on the left is for electrons, the one on the
right is for protons (as labelled).
We assume $t=10$ and $\theta=\pi/2$.
} 
\label{bgamma}
\end{figure}

One can ask if $\gamma\sim$20 and $B\sim 10$ G are
indeed required in order to have a negative cross section in the GHz band.
To this end, in Fig. \ref{bgamma}, we show the few lines corresponding to
two requirements:
\begin{enumerate}
\item the cross section must turn negative in the GHz band.
Taking the expansion of  $d\sigma_{\rm s}/ d\Omega$   (Eq. \ref{ds}) for $y\gg 1$
(see Eq. 2.12b of GS91),  one has that $d\sigma_{\rm s}/ d\Omega$
turns negative for
\begin{equation}
t \, >\, 2 +{87\over 30 y}  +O(y^{-2}) \to y > {87\over 30 (t-2)}
\end{equation}
implying
\begin{equation}
\gamma \, < \, \left[ {10\over 87}\, {2\pi m_{\rm e}c  
\over e B\sin\theta} \, (t-2)(1+t)^{3/2}  \right]^{1/2}
\label{bg}
\end{equation}
\item the absolute value of the (negative) cross section must
be large.
For this we require the normalization of the cross section:
\begin{equation}
k_\sigma \equiv (2/9)(e/B) \gamma^{-4} \sin^{-2}\theta 
\label{ksigma}
\end{equation}
to be as large as possible. 
\end{enumerate}

Fig. \ref{bgamma} shows the region (hatched) where the cross section
turns negative (labelled ``electrons"), together with the lines 
of constant $k_\alpha$. 
We assume  $t=10$ and a pitch angle $\theta=\pi/2$.
The smaller $B$, the larger $k_\alpha$:
the values of $B\sim 10$--100 and $\gamma\sim 20$ are typical
for having an efficient electron synchrotron maser.

\subsection{Synchrotron masers by protons}

The normalization of the synchrotron cross section
is of the order of $e/B$, and does not depend on the mass of the
emitting particles.
The mass controls instead the range of frequencies where both
the emission and the absorption cross section operates.
On the other hand, this range becomes the same if 
$\gamma m /B$ is the same: protons with the same $\gamma$
of the electrons, but with a magnetic field $m_{\rm p}/m_{\rm e}$
larger, emit the same frequencies.

Therefore Fig. \ref{bgamma} shows how the two constraints discussed above
select another preferred region in the $\gamma$--$B$ plane when the
emitting particles are protons: $\gamma\sim 20$ and $B=10^4$--$10^5$ G.
In this case the normalization of the (negative) cross section is smaller
than in the electron case, since $k_\alpha\propto e/B$, and $B$ is greater.
 
\section{Discussion}

The synchrotron maser described above works if in the emitting region there is 
a high degree of order.
In particular, the distribution of pitch angles should be narrower than $1/\gamma$. 
Otherwise, there will be absorption angles smaller than $1/\gamma$, and the 
photon would be truly absorbed.
It is surely difficult to have such an anisotropic pitch angle distribution
even in a limited region of space, but one possibility might be to have a magnetic mirror. 
In fact the magnetic moment $\mu\propto (\sin^2\theta) / B $ 
is a constant of motion, and if the particles moves towards a region of a greater $B$,
it increases its pitch angle, and eventually it is bounced back. 
When this occurs, the pitch angle is $\pi/2$. 
Particles with initially different pitch angles will bounce in different locations,
very close to the start if their pitch angle is already close to $\pi/2$, and
far away if it is small.
In between, we will have the presence of particles with different pitch angles,
except in the far away zone, where we find only particles with $\theta=\pi/2$.
This also requires that there are no particles with initial very small pitch angles,
that never bounce.
This ``segregation" of particles can offer a way to have a region 
where particles have the same pitch angle of $\pi/2$, and where the
synchrotron maser can occur.
Convergent magnetic field lines are very common in astrophysical sources:
a dipole magnetic field around neutron stars and white dwarfs,
or a stellar protuberance are two examples.

Based on the results above, we can envisage two very different scenarios
for synchrotron maser to play a role.
First, if the emitting particles are electrons, we have seen that the 
favoured magnetic field is relatively small, of the order of 10--100 G.
This is what we have in normal stars. 
This would suggest a Galactic origin of FRBs.

On the other hand, if the emitting particles are protons, the likely
magnetic field is of the order of $10^4$--$10^5$ G. 
This is what one expects close to the surface of white dwarfs.
Also in this case a Galactic origin if preferred, because it is likely
that a white dwarfs cannot produce the energetics required by extragalactic FRBs.

The other possibility is to have neutron stars. 
At a distance of a few hundreds of neutron star radii we would have
(for a dipole $B\propto R^{-3}$ field), the right values
of magnetic field for a large cross section for stimulated emission.
This could be at or very close to the light cylinder.
According to the typical power released in these events, neutron stars could be in our Galaxy
or cosmological.

Let us estimates the required number density in different scenarios.
Let us assume that particles of mass $m$ (left unspecified, it can be $m_{\rm e}$ or $m_{\rm p}$)
of the same energy $\gamma mc^2$, with the same pitch angle $\theta=\pi/2$, 
occupy a localized and magnetized region of space of size $R$.
Let assume that the particle number density $n$ is constant throughout the region.
Although the main emission occurs within an angle $1/\gamma$, there will be some
photons emitted at an angle larger than $1/\gamma$. 
The number of these photons will be suddenly amplified by stimulated emission.
Soon, the density of these photons ($n_\gamma$) exceeds the density of particles, and
the exponential amplification stops. 
After this time, the increase in photon density is linear in time.
The total number of photons leaving the source as a response of this initial
trigger is of the order of $R^3 n$.
They are all in phase, and distributed in a plane 
perpendicular to the direction of their velocity within an accuracy of $1/\gamma$.
In this case, the observed time interval is not directly related to the light 
crossing time, but rather to the duration of the maser, namely the duration
of the ``injection seed" photons or the timescale for which the emitting particles loose
a sizeable fraction of their energy.
In the latter case we have:
\begin{equation}
n  R^3 \gamma m c^2 \, =\, {E_{\rm FRB}\over \gamma^2} \, \to \,
n  \, =\, {E_{\rm FRB}\over \gamma^3R^3 m c^2}
\label{enne}
\end{equation}
where $E_{\rm FRB} = 10^{-3} \Delta t_{\rm ms } L_{\rm FRB}$ and the factor $1/\gamma^2$ accounts for 
the collimation of the observed radiation.

%
%

If FRBs are associated to stellar flares, then the typical size should be $R\sim 10^9$--$10^{10}$ cm,
the magnetic field of the order of 10--100 G,
implying that the emitting particles are electrons.
Their typical density should be $n = n_{\rm e} \approx 10^3 E_{\rm FRB,\, 27}/(\gamma_1 R_9)^3$ cm$^{-3}$.
We obtain a density $\sim 10^3$ smaller in the case of a Galactic white dwarf, with $B\sim 10^4$ G and 
protons as emitting particles.
For a neutron star, $B\sim 10^4$ G is the (dipole) field at a few hundreds of star radii,
therefore we have again $R\sim 10^9$ cm.
This yields $n\approx 70 \, E_{\rm FRB,\, 27}/(\gamma_1^3 R_9^3)$ cm$^{-3}$ 
if FRBs are Galactic.
These densities are small enough to fulfil the constraints on the frequency dependence
of the dispersion (see Eq. 4 of Katz 2016c) and the transparency (Eq. 5 of Katz 2016c).

If FRBs are cosmological, the known sources that can produce energetic events with 
$E_{\rm FRB}\sim 10^{39}$ erg (excluding GRBs) are neutron stars, magnetars 
and Active Galactic Nuclei.
In these sources the proton synchrotron maser is a viable option,
with magnetic fields of the order of $B\sim 10^4$--$10^5$ G.
For extragalactic neutron stars, the density estimate given above now gives
$n\approx 7\times 10^{13} E_{\rm FRB,\, 39}/(\gamma_1^3 R_9^3)$ cm$^{-3}$.
This is a rather large density, and a potential problem, since
a cloud with similar densities, surrounding the source, makes the frequency
dependence to largely deviate from the $\Delta t \propto \nu^{-2}$ observed law.

These are, admittedly, very rough estimates.
On the other hand, the purpose of this paper is to indicate a novel emission mechanism 
to obtain coherent radiation with extremely large brightness temperatures. 
The synchrotron maser has the advantage to avoid the problems associated to particle bunching,
and greatly relaxes the problems to explain the very short duration of the observed pulses,
because they are no longer simply associated with the size of the emitting region.

In a forthcoming paper I plan to explore the possibility to have 
masers from the curvature radiation process, to extend the applicability
of radio masers not only to FRBs, but also to radio pulsars.

\section*{Acknowledgements}
I thank Sergio Campana, Fabrizio Tavecchio and Giancarlo Ghirlanda for discussions.



\end{document}